\documentclass[sigconf]{acmart}
\renewcommand\footnotetextcopyrightpermission[1]{} %
\usepackage{amsmath,amssymb,amsfonts}
\usepackage{algorithm}
\usepackage[noend]{algpseudocode}
\usepackage{graphicx}
\usepackage{textcomp}
\usepackage{xcolor}
\usepackage{xspace}
\usepackage{url}

\usepackage{paralist}

\usepackage{booktabs}

\definecolor{pblue}{rgb}{0.13,0.13,1}
\definecolor{pgreen}{rgb}{0,0.5,0}
\definecolor{pred}{rgb}{0.9,0,0}
\definecolor{pgrey}{rgb}{0.46,0.45,0.48}
\definecolor{java}{rgb}{0.2, 0.4, 0.9}
\usepackage{listings}
\begin{document}

\newcommand{\eot}{\texttt{</t>}\xspace}
\newcommand{\id}[1]{\textsf{\footnotesize #1}}
\algrenewcommand\algorithmicindent{0.75em}%

\title{Maybe Deep Neural Networks are the Best Choice for Modeling Source Code\\}

\author{Rafael-Michael Karampatsis}
\affiliation{
  \institution{The University of Edinburgh}
  \city{Edinburgh}
  \country{United Kingdom}
}

\author{Charles Sutton}
\affiliation{
  \institution{Google AI, The University of Edinburgh and The Alan Turing Institute}
}

\thanks{This work was supported in part by the EPSRC Centre for Doctoral Training in Data Science, funded by the UK Engineering and Physical Sciences Research Council (grant EP/L016427/1) and the University of Edinburgh.
}

\newcommand{\maybecomment}[1]{{\lbrack #1 \rbrack}}
\newcommand{\todo}[1]{{\color{red}\maybecomment{TODO: #1}}}
\newcommand{\cs}[1]{{\color{purple}\maybecomment{From Charles: #1}}}
\newcommand{\raf}[1]{{\color{blue}\maybecomment{From Rafa: #1}}}

\newcommand{\ra}[1]{\renewcommand{\arraystretch}{#1}}

\begin{abstract}
Statistical language modeling techniques 
have successfully been applied to source code, yielding
a variety of new software development tools, such as tools for code suggestion
and improving readability.
A major issue with these techniques is that code introduces new vocabulary
at a far higher rate than natural language, as new identifier names proliferate.
But traditional language models limit the vocabulary to a fixed set of common words.
For code, this strong assumption has been shown to have a significant negative effect on predictive performance. 
 But the open vocabulary version of the neural network language
models for code have not been introduced in the literature.
We present a new open-vocabulary neural language model for code 
that is not limited to a fixed vocabulary of identifier names.
We employ a segmentation into subword units, subsequences of tokens chosen 
based on a compression criterion, following previous work in  machine translation.
Our network achieves best in class performance, outperforming even the state-of-the-art methods of  Hellendoorn and Devanbu that are designed specifically
to model code.
Furthermore, we present a simple method for dynamically adapting the model
to a new test project, resulting in increased performance.
We showcase our methodology on code corpora in three different languages
of over a billion tokens each, hundreds of times larger than in previous work.
To our knowledge, this is the largest neural language model for code 
that has been reported.
\end{abstract}



\maketitle

\section{Introduction}

Large corpora of open source software projects present an opportunity
for creating new software development tools based on machine learning
\cite{big-code-survey}. Many of these methods are based on the hypothesis
that much of software is \emph{natural}, that is, because software is written 
for humans to read, it displays some of the same statistical properties
as natural language.
To quantify the degree of naturalness of a piece of software, Hindle et al \cite{Hindle2012} propose the use of statistical language modeling.
A language model is a probability distribution over strings;
by training a language model (LM) on a large corpus of well-written code, we hope that the LM will assign high
probability to new code that is similar to the training
set,  in other words, code that is
 well-written, easy to read, and
natural.
There is now a large literature on language modeling for code
\cite{Hindle2012,Allamanis2013,Nguyen2013,Tu2014,Bielik2016,Dam2016,Hellendoorn2017}.

Such models have enabled research on a broad suite of new software engineering tools.
For example,
the ability to automatically quantify the  naturalness
of software has enabled 
new tools for autocompletion \cite{Hindle2012,Raychev2014}, improving code readability \cite{Allamanis2014,raychev15}, and program repair \cite{Ray2016}.
Furthermore, recent work in natural langugage processing (NLP) \cite{Peters2018, Devlin2018} has shown that 
LMs and sequence models in general learn useful word embeddings, which can then be used
for downstream tasks in the same way as older, 
word2vec-style embeddings \cite{Mikolov2013}.
Such continuous embeddings have formed the foundation of important software engineering tools.
Examples of such are suggesting readable function and class names \cite{Allamanis2015a}, summarizing source code \cite{Iyer2016, Allamanis2016}, predicting bugs \cite{Pradel2018}, detecting code clones \cite{White2016}, comment generation \cite{Hu2018}, fixing syntactic errors \cite{Kruthiventi2015}, and variable de-obfuscation \cite{Bavishi2018}.
Therefore, improved LMs for code have the potential to enable improvements in a diverse
variety of software engineering tools.

However, a provocative recent paper by Hellendoorn and Devanbu \cite{Hellendoorn2017} argues
that there is a key general challenge in deep learning models of code, which significantly
hinders their usefulness.
This is the \emph{out of vocabulary ({OOV})} problem, which is
that new identifier names are continuously invented by developers \cite{Allamanis2013}. 
These \emph{out-of-vocabulary (OOV)} tokens cannot be predicted by a language model with a fixed vocabulary,
because they have not occurred in the training set.
Although this problem also exists in natural languge text, it is much more severe in code.
For instance, the vocabulary size of 74046 tokens used in \cite{Hellendoorn2017} covers only about 85\% of identifier occurrences and 20\% of distinct identifiers that appear in the test set.
Therefore
standard methods from NLP of dealing with this problem, such as
introducing a single special token to represent OOV tokens \cite{Chen1996}, still fall short of 
fully addressing the OOV problem for code.
Instead, Hellendoorn and Devanbu \cite{Hellendoorn2017} present a $n$-gram
LM with several code-specific extensions. This enhanced
$n$-gram model shows
improved performance over
an off-the-shelf neural language model for code, a surprising result
because for natural language, neural models consistently outperform
$n$-gram models.
Based on these improvements, Hellendoorn and Devanbu raise the provocative
suggestion that deep models might not be the best choice for modeling sournce
code.

In this paper, we argue that to the contrary, perhaps deep networks \emph{are} a good
choice for modeling source code, because it is possible
to overcome the limitations highlighted by \cite{Hellendoorn2017}.
More specifically, we address the key challenge of deep models of source code
that was highlighted in the previous work, namely, the OOV problem, 
by introducing an open-vocabulary neural language model
for source code.
An \emph{open vocabulary} model is not restricted to a fixed-sized vocabulary determined at training time;
for example, some types of open vocabulary models predict
novel tokens character-by-character.
Our open vocabulary
model is based on the idea of \emph{subword units}, following previous work from neural machine translation \cite{Sennrich2015}.
Subword units are a way 
of combining the strengths of character-level and token-level models. Each subword unit is a sequence of characters that occurs as a subsequence of some token in 
the training set; the model outputs a sequence of subword
units instead of a sequence of tokens.
Including all single characters as subword units will allow
the model to predict
all possible tokens, so there is no need for special
OOV handling.
The vocabulary of subword units in the model is inferred from the training set using a compression-based
heuristic called byte pair encoding (BPE).

The use of subword unit NLMs has two main advantages for code: 
First, even for an OOV token $t$ that has never occurred in the training data, its subword units will have occurred in training, so the model can predict what will follow $t$ based on what tended to follow its subword units
in the training data.
Second, training of a subword unit NLM on large corpora corpora is much faster than a token level model, as
a relatively smaller number of subword units can
still lead to good performance.
On large open-source corpora, we show that our model is indeed
more effective
for code language modeling than previous $n$-gram or neural models.
On Java, our model is able to achieve a predictive performance of
3.15 bits per token and 70.84\% MRR in a cross project
evaluation. Simultaneously achieving predictive performance of \emph{1.04 bits per token} and 81.16\% MRR in a within project setting,
This is the best performance that we are aware of in the literature for a single model.

In particular,
our contributions can be summarised as follows:
\begin{itemize}
\item We present the first open-vocabulary neural language model (NLM) for source code.
This is based on an automatic segmentation of code tokens into smaller subword units that 
learns the common statistical internal patterns within
identifier names.
\item We show that our model outperforms previous
neural and $n$-gram LMs for code across three programming languages: Java, C, and Python. For C and Python, 
we are the first to showcase deep learning results for corpora of our size.  Interestingly, we show that while
the $n$-gram LMs are unable to improve their performance
on larger training sets, neural LMs are able to improve
their performance given more data.
\item  Ours is the largest neural LM for code
reported in the literature, trained on 1.7 billion tokens, which is 107 times larger than in previous work.
\item We present a simple heuristic to adapt a NLM trained
on a large corpus of projects to a single project of interest, an important scenario for practical development
tools.
This allows us to report results on the code maintenance scenario of \cite{Hellendoorn2017}, which was previously infeasible for NLMs.
\end{itemize}

\section{Related work}
\label{sec:related}

\paragraph{Language modeling for code}
Numerous researchers have applied language modeling techniques on code corpora.
These studies are based on the assumption that software is characterized by similar statistical properties to natural language; viz., the naturalness hypothesis \cite{Hindle2012}.
It should be expected that software is repetitive and predictable.
Indeed this successfully verified in \cite{Gabel2010}.
Hindle et al. \cite{Hindle2012} introduced the naturalness hypothesis and showcased that Java code is actually less entropic than a natural language (English). 
Nguyen et al. \cite{Nguyen2013} augmented $n$-gram LMs with semantic information such as the role of a token in the program, e.g., variable, operator, etc.
On the other hand, other researchers attempted to exploit the localness of code by augmenting $n$-gram models with a cache component \cite{Tu2014}. The cache contains 
$n$-grams that previously appeared in the current file or in nearby ones,
such as files from the same package. The model was shown to offer lower entropies for code but not for English.
Later, Hellendoorn and Devanbu \cite{Hellendoorn2017} extended this idea to nested scopes, outperforming vanilla NLMs and achieving best-in-class performance on a Java corpus.

Beyond $n$-gram models, several other types of methods have been employed to model code.
A generative model for code called probabilistic higher order grammar (PHOG) was introduced by \cite{Bielik2016}, which generalizes  probabilistic context free grammars \cite{Jurafsky2000}.
Also, both simple RNNs were used in \cite{White2015} and LSTMs \cite{Dam2016} to learn an NLM for code. The LSTMs were shown to perform much better than simple RNNs. This conclusion was later confirmed by Hellendoorn and Devanbu \cite{Hellendoorn2017}.
Lastly, \cite{Li2018} attempt to improve code completion performance on OOV words by augmenting an RNN with a pointer network \cite{Vinyals2015}.
Their pointer network learns a copy mechanism over the current context using attention that is useful in code completion. 
Once an OOV token has been used once, the copy mechanism can learn to re-use it, but unlike our model it cannot predict its first usage and it is not designed to learn dependencies between the OOV token and the next tokens as it learns no representations for OOV words, in contrast to our method.
That said, the subword units learned by our model can be used by any sort of neural model for code,
so it could be fruitful in future work to combine our approach with this previous work,
for example by augmenting our network with a pointer network component.

\paragraph{Applications of code language models}
Probabilistic code models have enabled many applications in software engineering.
One example is recommender systems aiming to aid developers in writing or maintaining code \cite{Mens2014}.
Hindle et al. used a token-level LM for code completion \cite{Hindle2012}, 
while later, Franks et al. \cite{Franks15} improved on performance using the cache n-gram from \cite{Tu2014} and built a code suggestion tool for Eclipse \cite{Gamma2003}.
Another application are recommendation systems for variable, method, and class names suggestion \cite{Allamanis2014, Allamanis2015a, Allamanis2016} that employ relevant code tokens as the LM context.
Campbell et al. \cite{Campbell2014} used $n$-gram language models to detect syntax error locations in Java code.
Lastly, Ray et al. \cite{Ray2016} showcased that buggy code has on average lower probability than correct one and that LMs can spot defects as effectively as popular tools like FindBugs. 

\paragraph{Out-of-vocabulary problem}
Even after an LM has been trained on a large corpus of projects,
many identifier names are still encountered which are out of vocabulary (OOV) token that have not been previously encountered in the training set.
These are called \textit{neologisms} by \cite{Allamanis2015a}.
Indeed on our corpora we find that OOV words are many times more common
in code than in natural language.
Traditional LMs are closed vocabulary, meaning that the vocabulary 
is fixed based on a training corpus, and they are unable to predict OOV words,
which is an obvious issue for code.
Apart from that, many identifiers appearing in the training set are rare.
This sparsity could potentially confuse the model or result in slower estimation of parameters.

To combat this problem,
previous research in language modeling for code has segmented identifiers via a heuristic \cite{Allamanis2015a}, which splits them on camel case and underscores.
Even though the resulting segmentation has the ability to handle some neologisms, it is limited to only combinations of subtokens appearing in the training set and thus unable to achieve an open vocabulary.
Additionally, many of these subtokens are still infrequent, which hinders the model's ability to assign high scores to their compositions.
A separate line of work is that several studies have empirically compared different 
techniques for automatically splitting identifiers \cite{enslen2009mining,hill2014empirical}. However, this work considers
a different problem than us. That previous work
focuses on splitting identifiers into words in a way that matches human
judgements. Although our approach also performs identifier splitting,
our goal is to split identifiers in a way that improves a language model;
in many cases, our subword units will be sequences of characters that are
not words.
In our application, 
it is of no interest whether our subword units correspond to words that humans recognize,
because our subword units can be trivially reassembled into complete
tokens before they are shown to a developer.

\paragraph{Open vocabulary language models}
The occurrence of OOV entries in test data is not a code specific problem but has also been an in issue in NLP, especially in
morphologically-rich languages, for example.
Alternative models have been proposed where the vocabulary is open.
Character language models are one such solution where each word is represented as a sequence of its characters.
While  recurrent NLMs have shown excellent performances for  character-level language modeling \cite{Sutskever2011, Hermans2013}, the performance of such models is usually worse than those built on the word level \cite{Mikolov2012}.
In order for character model to capture the same dependencies as a token level one, the length of the sequences that gradients are calculated upon needs to be increased. This can cause training to become more difficult due to the
 vanishing or exploding gradient problems, even for LSTM \cite{Hochreiter1997} and GRU \cite{Cho2014a} recurrent networks which are designed specifically to address gradient problems. 

Another option that attempts to open the vocabulary is to represent words as a sequence of segments that when merged result to the original word.
For example, language models have been proposed at the morpheme level\cite{Luong2013}, the phone level \cite{Bazzi2002},
and the syllable level \cite{Mikolov2012}.
Other models combine character level models with a caching mechanism
to reuse generated tokens 
\cite{kawakami2017learning,chung2016hierarchical}.
Finally, other researchers have attempted to also learn the segmentation  on a text corpus for machine translation \cite{Sennrich2015}, which is the approach
that we build on in our work.
Subword unit segmentations have been used to capture morphology \cite{Vania2017} in a language modeling task but the network's output were on the word level and thus unable output OOV entries.
Simultaneously with our work, \cite{Mielke2018} independently developed a recent LM based on subword units for natural language, and found that it had close to state-of-the-art performance.

\section{Methodology}
\label{sec:methodology}
In this section, we present our neural LM for code based 
on subword units. We begin by giving a brief background
on the neural language models we use (Section~\ref{sec:rnnlm}). Then we describe how 
we construct the vocabulary of subword units (Section~\ref{sec:bpe}), followed by a search
procedure for producing $k$-best completions (Section~\ref{sec:beam-search}), 
and finally we discuss how we adapt the model on a new project (Section~\ref{sec:adaptation}).

\subsection{Neural Language Model with GRU Cell}
\label{sec:rnnlm}

State-of-the-art LMs for natural language
are currently based on recurrent neural networks (RNNs)
\cite{Mikolov2010, Sundermeyer2015,mellis17lstm}.
RNN language models scan an input sequence forward 
one token at a time, predicting a distribution over each
token given all of the previous ones.
RNNs with gated units, such as long short-term memory (LSTM) units \cite{Hochreiter1997} 
and gated recurrent units (GRUs) \cite{Cho2014b}, have
been found to outperform other methods for language modeling \cite{Sundermeyer2015,Dam2016}.
Intuitively, the advantage of an RNN over older language models, such as $n$-gram language models, is that an $n$-gram model uses only a short window of $n$ tokens to predict the
next token, whereas an RNN can potentially take into account the entire previous
history of the sequence.
The advantage of gated units are that the gates allow the network to learn when to forget information from the hidden state and take newer, more important information into account \cite{Hochreiter1997}.
Among different kinds of gated units, 
GRUs have been shown to perform comparably to LSTMs across different applications \cite{Chung2014}.
In our initial experiments we found GRUs to slightly outperform LSTMs when trained on the Java corpus,
so we use them in our model.

Our model is a single layer GRU NLM built upon subword units which have been learned from BPE as described in Section~\ref{sec:bpe}.
For each vocabulary entry we learn a continuous representation of 512 features, while the GRU state is of the same size.
In all our experiments we used a learning rate of 0.1, dropout of 0.5  \cite{Srivastava14} and a maximum of 50 training iterations using stochastic 
gradient descent with a minibatch of 32 for the small training sets and a minibatch size of 64 for the full training sets.
After each iteration we tested the network on a validation set and measured its cross entropy (see Section~\ref{sec:intrinsic}).
If the cross entropy is larger than the previous epoch then we halve the learning rate and this can happen for a maximum of 4 times, otherwise training stops.
During training of the global model we unroll the GRU for 200 timesteps.
Our implementation is open source, written in Tensorflow \cite{Tensorflow2015} and it is available in a public GitHub repository.\footnote{\url{https://github.com/mast-group/OpenVocabCodeNLM}}

\subsection{Selecting Subword Units Using Byte Pair Encoding}
\label{sec:bpe}

Traditional language models in NLP most commonly operate at the token level \cite{Sundermeyer2015,Dam2016}, meaning that the RNN predicts one token at a time.
But for code, this strategy leads to large vocabulary sizes, because identifiers
in programming languages often correspond to entire phrases in natural language.
Because the number of unique identifiers increases with the size of the corpus
\cite{Allamanis2013},
this problem makes it infeasible to train code LMs on large corpora.
As we later illustrate in Section~\ref{sec:rq2Definition} the vocabulary of three different giga-token code corpora is an order larger than an equivalent English one.

In our code LM, we address this problem by having the model predict \emph{subword units} rather than full tokens at each time
step of the RNN. A subword unit is an $n$-gram of characters
that appear as a subsequence of some token in the corpus. An example of a Java source file
segmented into subword units is shown in Figure~\ref{fig:JavaCodeBPETokens}.
Notice that we include a special subword unit \eot that marks the end of a token, allowing
us to convert from a sequence of subword units back into a sequence of tokens.
The subword units in the model are chosen adaptively
based on statistical frequency, as we will describe shortly.
The effect of this is that more common tokens, like \texttt{public} in  Figure~\ref{fig:JavaCodeBPETokens} are assigned a full subword unit,
whereas less common tokens, like \texttt{setter}, are divided into
smaller units that are individually more common.

\begin{figure}
\footnotesize
{ \center Java Code:} \vspace{0em}\\
\begin{lstlisting}[basicstyle=\footnotesize]
public AttributeContext(Method setter, Object value)
{
   this.value = value;
   this.setter = setter;
}
\end{lstlisting}

{\center Subword Units:} \vspace{0.5em}\\
\tt\raggedright
public\eot\ \ Attribute\ \ Context\eot\ \ (\eot\ \ Method\eot\ \ set\ \ ter\eot\ \ ,\eot\ \ Object\eot\ \ value\eot\ \ )\eot\ \{\eot\ \ this\eot\ \ .\eot\ \ value\eot\ \ =\eot\ \ value\eot\ \ ;\eot\ \ this\eot\ \ .\eot\ \ set\ \ ter\eot\ \ =\eot\ \ set\ \ ter\eot\ \ ;\eot\ \ \}\eot
\caption{Subword units list for a Java function.}
\label{fig:JavaCodeBPETokens}
\end{figure}

The use of a subword unit LM for code has two potential advantages.
First, because the model has a smaller vocabulary size, 
it may have better performance because of a reduced level of data sparsity.
Second, the model can synthesize OOV tokens that have not been seen in the training
data via the smaller subtoken units.
The vocabulary of subword units is learned before training the NLM by segmenting 
a corpus of code. This is done in such a way that more frequent character $n$-grams are 
more likely to be included in the vocabulary of subwords units.
This strategy results in a core vocabulary of subword units that occurs frequently
across different projects and captures statistical patterns of characters within 
identifiers.

In order to learn the segmentation we use a modification of \emph{byte pair encoding (BPE)} \cite{Gage1994}.
BPE is a data compression algorithm that iteratively finds the most frequent pair of bytes in the vocabulary appearing in a given sequence, and then replaces it with a new unused entry.
Sennrich, Haddow, and Birch \cite{Sennrich2015} first adapted the algorithm for word segmentation so that instead of merging pairs of bytes, it merges pairs of characters or character sequences.
The learned segmentation was used in their neural translation system and resulted in improved translation of rare words.

The algorithm starts with a vocabulary containing all single characters in the data set plus the \eot symbol.
All symbol pairs appearing in the vocabulary are counted and we then replace all the appearances of the most frequent pair $(S_1, S_2)$ with 
a unique new single symbol $S_1S_2$, which we also add to the vocabulary.
This procedure is called a merge operation $(S_1, S_2) \rightarrow S_1S_2$. 
The algorithm stops after a given maximum number of merge operations to be performed is reached.
We clarify that as in \cite{Sennrich2015} we do not consider merging pairs that cross token boundaries, that is, where the merged token would contain
\eot internally, so that every subword unit is a character subsequence of
a token in the data.
The final output of the algorithm is the new vocabulary, which
contains all the initial characters plus the symbols created from the merge operations, and the ordered list of merge
operations performed in each iteration.
We run the BPE algorithm on a held out dataset of projects
that are separate from the training, validation, and test sets.
We experimented with three different encoding sizes, i.e., the maximum
number of merge operations: 2000, 5000, and 10000 operations.

To train the LM, we first segment the train, validation, and test sets using the learned encoding. To do this,  
we transform each token into a sequence of its characters, adding
\eot symbols after every token. Then we apply in order the merge operations 
from BPE to merge the characters into subword units in the vocabulary.\footnote{The BPE implementation that we used was taken from \url{https://github.com/rsennrich/subword-nmt}} Finally, we train and test a GRU LM in
the usual way on the data that has been segmented into subword units.

\subsection{Predicting Best $k$ Tokens}
\label{sec:beam-search}

In an autocompletion setting, it might be desirable
to present a ranked list of $k$ predicted tokens rather than
a single best prediction.
But because our model is based on subword units, it is not completely trivial to generate top $k$ predictions
of full tokens, because a single token could be made from many subword units.
We approximate these using a beam-search-like algorithm.
If the beam is large enough the algorithm can give a good approximation of 
the top-$k$ complete tokens.

More specifically, the NLM defines a probability $p(s_1 \ldots s_N)$
for any sequence of subword units. The goal of the search procedure
is: given a history $s_1 \ldots s_N$ of subword units that already appear
in a source file, predict the complete token that is most likely to occur next. A \emph{complete token} is a sequence of subword units 
$w_1 \ldots w_M$ that comprise exactly one token: that is, $w_M$ ends
with \eot and none of the earlier subword units do.
The goal of the search algorithm is to find the $k$ highest probability complete tokens, where we denote a single token as the sequence of units
$w_1 \ldots w_M$,
that maximize the model's probability
$p(w_1 \ldots w_M | s_1 \ldots s_N)$. Importantly, the length $M$ of the new complete
token is \emph{not} fixed in advance, but the goal is to search over complete
tokens of different length.

The algorithm is illustrated in Algorithm~\ref{alg:Decoder}. 
Given a value of $k$ and a beam size $b$, it starts by querying the model to obtain
its predictions of  possible subword units, ranked by probability;
in our pseudocode, we assume that the model's \id{predict} function returns
a ranked list of a given size, and that $V$ is the total size
of the vocabulary.
The algorithm uses two priority queues: one called \id{candidates} 
which ranks the sequences of subword units that still need to be
explored during the search, 
and one called \id{bestTokens} which contains the $k$ highest
probability complete tokens that have been expanded so far.
Each candidate is a structure with two fields, \id{text} which
is the concatenation of all the subword units in the candidate,
and \id{prob} which is the product of the probabilities of each subword unit
in the candidate. The candidate class has an \id{extend} method
which updates both of these fields in order to add one additional
subword unit to the end of the candidate. Both of the priority queues
are sorted by the probability of the candidate.

The main loop of the search is in lines \ref{alg:termination}-\ref{alg:endSearch}.
In each iteration, the algorithm pops the $b$ best candidates from the 
\id{candidates} queue, expanding them and scoring their expansions, in which each candidate
is extended by one additional subword unit.
If an expansion creates a token, that is, the new subword unit
ends with \eot, then it is pushed onto the token queue and the worst token is popped.
This maintains the invariant that \id{bestTokens} has size $k$.
If the new expansion is not a complete token, then it is pushed onto
the \id{candidates} queue, where it can potentially be expanded in the next iteration.

This search procedure is repeated until any of the following termination criteria has been satisfied at line \ref{alg:termination}:
\begin{enumerate}[(a)]
\item The number of complete tokens that have been explored during the search exceeds
a threshold (in our implementation, we use $\id{tokensDone} > 5000$).
\item The cumulative probability of all the tokens that have 
been explored exceeds the threshold, i.e. $\id{total} > 0.8$
\item A sufficient number of 
search iterations have been completed, i.e. $\id{iters} > 7$.
\item The probability of the best candidate is less than the worst current complete top-$k$ tokens, that is,
$$\min \{ c.\id{prob} \,|\, c \in \id{bestTokens} \}  \geq 
\max   \{ c.\id{prob}\,|\,c \in \id{candidates}) \}.$$
Expanding a candidate cannot increase its probability, so at this point we are guaranteed that no better complete tokens will be found in the remainder of the
search.
\end{enumerate}
These criteria ensure that the beam search always terminates.

\begin{algorithm}
\caption{Predicts top $k$ most likely tokens according to the model to follow the history
of subword units $s_1 \ldots s_N$.} \label{alg:Decoder}
\begin{algorithmic}[1]
\small
\Procedure{PredictTopK}{$\id{model}$, $s_1 \ldots s_N$, $k$, $b$, $V$}
    \State {\id{subwords}, \id{probs} $\gets$ \id{model}.\id{predict}$(V, s_1 \ldots s_N)$}
    
    \vspace{1ex}\State \text{\# \sl Initialize priority queues of completed tokens}
    \State $\id{bestTokens}\gets \text{$k$ highest probability tokens from \id{subwords}}$
    \State $\id{candidates} \gets \text{$b$ highest probability non-tokens from \id{subwords}}$
    \State $\id{total} \gets \id{sum}(\text{$c$.\id{prob} for $c$ in \id{bestTokens}})$

\vspace{1ex}\State \text{\# \sl Main search loop. Expand $b$ best incomplete tokens}

   	\State $\id{lowest} \gets \id{min}(\text{\id{t}.\id{prob} for $t$ in \id{bestTokens}})$
   	\While{termination criterion not met}\label{alg:termination}
    	\State $\id{toExpand} \gets \id{candidates}.\id{popNBest}(b)$
    	\ForAll{$\id{candidate} \in \id{toExpand}$}
    		\State \id{subwords}, \id{probs} $\gets$ \id{model}.\id{predict}$(b, \id{candidate.text})$
    		\ForAll{$w \in \id{subwords}$}
		        \State $\id{newCandidate} \gets \id{candidate}.\id{extend}(w, \id{probs}[w])$
    			\If{$\id{isToken}(\id{newCandidate})$}
    				\State $\id{bestTokens.push}(\id{newCandidate})$
    				\State $\id{bestTokens.pop}()$ \text{\hspace{5em}\# \sl Retain top $k$}
    				\State $\id{lowest} \gets \id{min}(\text{\id{t}.\id{prob} for $t$ in \id{bestTokens}})$
    				\State $\id{total} \gets \id{total} + \id{newCandidate}.\id{prob}$
    				\State $\id{tokensDone} \gets \id{tokensDone} + 1$
    			\Else
    				\State $\id{candidates}.\id{add}(\id{newCandidate})$
    			\EndIf
    		\EndFor
    	\EndFor
    \State $\id{iters} \gets \id{iters} + 1$
    \EndWhile\label{alg:endSearch}
    \State \Return \id{bestTokens}
\EndProcedure
\end{algorithmic}
\end{algorithm}

\subsection{Dynamic adaptation to new projects}
\label{sec:adaptation}

It is important to be able to quickly adapt a \emph{global LM}, which has
been trained on a diverse corpus of projects, to have better
performance on a new project of interest. We call this \emph{dynamic adaptation}.
For example, suppose that an organization distributes an IDE that contains a code LM trained on a large number of Github projects, and a separate company is using the IDE for a new project.
The LM will be expected to perform worse on the new project \cite{Hindle2012}, so for performance it would be desirable to retrain the model on the new project.
However, for confidentiality reasons, the company may well be unwilling to send their code back to the IDE vendor for retraining the model.
Although a within-project model might be an alternative, a single project 
does not provide much training data for an NLM, especially in the early
stages of a project.
Therefore, it would be desirable to have a fast way of adapting the LM to a new project, in a way that requires access only to the trained LM and the source code of the new project.
In principle, we could train from scratch a new model on both the original training set and the new  project, but this would
be computationally expensive.

Instead, we use a simple method of dynamically adapting our global neural LMs to a new project. Given a new project, we start with the global LM and update
the model parameters by taking a single gradient step on each encountered sequence in the project after testing on it. This series of updates is equivalent to a single training epoch on the new project. (In our evaluations in Section~\ref{sec:evaluation}, we will split up the project files in such a way that we are never training
on our test set.) We unroll the GRU for
20 time steps instead of 200 as in our global models, in order to update the parameters more frequently.
Our choice of applying only one update is motivated by the following reasons.
First, it is faster, allowing 
the model to quickly adapt to new identifiers in the project.
Second, taking too many gradient steps over the new project could cause
the LM to give too much weight to the new project, losing information 
about the large training set of the global model.

\section{Datasets}
\label{sec:datasets}

In our experiments we used code corpora from three popular programming languages:
 Java, C, and Python. Although these languages are related, they also have
 differences that might be hypothesized to affect the performance of LMs.
Java was an obvious choice since it has extensively been used in related work \cite{Hindle2012,Allamanis2013,Nguyen2013,Tu2014,Dam2016,Hellendoorn2017}.
Unlike Java, C is not object oriented, and the language makes it possible to write exceptionally terse
code.\footnote{For examples, see \url{https://www.ioccc.org/}.} Finally,
Python is also object oriented but it is mainly a dynamic language with little
use of static typing. These differences between C and Python make them interesting to consider 
alongside Java.

For Java we used the Java Github corpus of Allamanis et al. \cite{Allamanis2013}, which consists of more than 14000 popular open source Java projects.
Following the procedure described in \cite{Allamanis2013}, the C corpus was mined in \cite{Dudoladov2013} and the Python corpus was mined in \cite{Fiott2015}.
For lexical analysis in Java we used the lexer implemented in \cite{Hellendoorn2017}%
\footnote{https://github.com/SLP-team/SLP-Core}, while C and Python code lexical analysis was performed via the Pygments\footnote{http://pygments.org/docs/lexers/} library. In Python, we do not add any special tokens to represent whitespace.
For all three languages, we preprocessed the data by replacing 
occurrences of non-ASCII character sequences 
such as Chinese ideograms inside strings with a special token that did not
occur elsewhere in the corpus. 

For Python and C we sampled 1\% of the corpus for validation and 1\% for testing. 
Another 10\% of the corpus was sampled as a separate data set upon which BPE was run to learn a subword encoding.
The rest of the data was used for training. We also report results on a smaller subset
of 2\% of our full training set.
For Java, we used a slightly different procedure to make our experiment comparable to a previous study \cite{Hellendoorn2017}. We divide the data into five subsets as in the other two languages. The validation and test sets are the same as in \cite{Hellendoorn2017}, and our ``small train'' set is the same as their training set. 
To obtain the full Java train set, we collect all of the files in the 
Java Github corpus that do not occur in the validation or test set.
Of these, we sampled 1000 random projects for the subword encoding data set,
and the remaining projects were used as the full train set.

\begin{table*}
\caption{Corpus statistics for each code corpus. 
}
\centering
\ra{1.3}
\begin{tabular}{@{}rrrcrrcrr@{}}
\toprule & \multicolumn{2}{c}{\textbf{Java}} & \phantom{abc}& \multicolumn{2}{c}{\textbf{C}} & \phantom{abc}& \multicolumn{2}{c}{\textbf{Python}} \\
\cmidrule{2-3} \cmidrule{5-6} \cmidrule{8-9} & Tokens & Projects && Tokens & Projects && Tokens & Projects \\ \midrule
Full\ Train       & 1436.39M & 13362 && 1685.71M & 4601 && 1056.33M & 27535 \\
Small\ Train      &   15.74M &   107 &&   37.64M & 177  &&   20.55M &   307 \\
Subword\ Encoding &   64.84M &  1000 &&  241.38M & 741  &&  124.32M &  2867 \\
Validation        &    3.83M &    36 &&   21.97M & 141  &&   14.65M &   520 \\
Test              &    5.33M &    38  &&  20.88M &  73  &&   14.42M &   190 \\
\hline 
\end{tabular}
\end{table*}

\section{Evaluation}

Our model was evaluated on both intrinsic and extrinsic evaluation measures.
An intrinsic metric judges the quality of an LM's 
predictions by themselves, 
in isolation from a larger task.
On the other hand, extrinsic methods perform indirect evaluation by assessing how a model affects performance of some other task.
We next describe the specific metrics used in our experiments.

\subsection{Intrinsic evaluation} \label{sec:intrinsic}
A good language model should assign high probability to a real sentence while simultaneously assigning a low probability to a wrong one.
In many applications in software engineering, accurate probability scoring is necessary, meaning that code fragments that are more likely to occur in human-written
code should be assigned higher probability.
Precise scoring of code fragments is essential for tasks like translating a program from one programming language to another \cite{Nguyen2013b, Karaivanov2014}, code completion \cite{Raychev2014, Franks15}, and code synthesis from natural language and vice versa \cite{Allamanis2015b, Desai2016, Nguyen2016, Raghothaman2016, Devlin2017, Bunel2018}.

The intrinsic metric that we use 
is cross entropy, which is a standard measure employed in previous work.
The cross entropy defines a score over a
a sequence of code tokens $t_1$, $t_2$, ..., $t_{|C|}$.
For each token $t_i$,
the probability of each token is estimated using the model under evaluation and it is denoted by $p(t_i|t_1,...,t_{i-1})$.
Then the average per token entropy is defined as:
\begin{equation}
H_p(C) = -\frac{1}{|C|}\sum_{i=1}^{|C|}\log p(t_i |  t_1,...,t_{i-1}).
\end{equation}
Cross entropy corresponds to the average number of bits required in every prediction.
Thus lower values are better.
This metric  not only takes into account whether
 the highest ranked prediction is correct, but also  
 rewards predictions with high confidence.

Our subword unit models define a distribution over subword units rather
than directly over tokens. To compute the cross entropy
for subword unit models, we segment each token $t_i$
into subword units $t_i = w_{i1} \ldots w_{iM}$. Then we compute the product
$$ p(t_i |  t_1,...,t_{i-1}) = \prod_{m=1}^M p( w_{im} |  t_1,...,t_{i-1},w_{i1} \ldots w_{i,m-1}),$$
where the right hand side can be computed directly by the subword unit NLM.
This probability allows us to compute the cross entropy $H_p(C)$.
(The technical reason that this method is correct is that discrete probability
distributions are preserved under 1:1 correspondences.)

\subsection{Extrinsic evaluation}

As an extrinsic performance measure, we report the performance
of our LMs on code completion, which is the task of
predicting each token in a test corpus given all of the previous tokens
in the file.
To measure performance on this task, we use
mean reciprocal rank (MRR).
MRR has previously been used in a plethora of code completion evaluations in relevant work \cite{Bruch2009, Raychev2014, Tu2014, Hellendoorn2017}.
The reciprocal rank of a query response is the multiplicative inverse of the rank of the first correct answer.
MRR is the average of reciprocal ranks or results for a sample of queries $Q$
defined as
\begin{equation}
\mathit{MRR} = \frac{1}{|Q|}\sum_{i=1}^{|Q|}\frac{1}{rank_{i}}.
\label{eq:MRR}
\end{equation}
For example, if the correct suggestion always occurs at rank 2 then the MRR 
is 0.5 roughly, at rank 10 the MRR is 0.1, 
and so on.
A simplified description of MRR is that it averages top-$k$ prediction accuracy across various $k$.
In this specific scenario $k \in [1,10]$ since the models output a list of top-$10$ best tokens.

\subsection{Test Scenarios} \label{sec:scenarios}
Our model was evaluated in three scenarios introduced in previous work \cite{Hellendoorn2017}, which they call
\emph{static}, \emph{dynamic}, and \emph{maintenance}
settings. Each setting simulates a different way of incorporating
code LMs within an IDE. For all settings, the task is to predict each
token in the test set,
but the training sets for each setting are slightly different.

\paragraph{Static tests} The model is first trained on a fixed training corpus, and is later evaluated on a separate test dataset.  Essentially this is a cross-project setting where the train, validation, and tests are all disjoint from each other and contain separate projects. This simulates the setting where a single
global LM is trained on a large corpus of projects and then deployed to
many different customers without any adaption.

\paragraph{Dynamic tests} In this setting, the model is allowed to update
its parameters after it has made predictions on files in the test set. 
In addition to the original training set, the model is allowed to retrain on 
files in the test set, after it has been scored on its predictions.
Note that the model is required to make its predictions on the testing file
before the file is added to the training set, so that we are never training on test data.
For our neural LMs, we adapt the model at test time using the procedure described in Section \ref{sec:adaptation}.
After we have finished evaluating each test project we restore the model to
the global cross-project one learned from the train set.
This simulates a setting in which some files are available from the test project
of interest for dynamic adaptation.

\paragraph{Software Maintenance tests} This scenario is perhaps the closest to real world usage. It simulates everyday development where a programmer makes small changes to existing code. In this setting, the LMs are tested on one file at a time in the test set. For each file, the full training set plus all other files in the test
 project apart from the file of interest is used as training data. 
Because this requires retraining the model once for each file in the test set,
this scenario was previously deemed infeasible for NLMs in \cite{Hellendoorn2017}.

\section{Research Questions}
\label{sec:evaluation}

When evaluating our models, we focus on the following research questions.

\emph{RQ1. How does the performance of subunit neural LMs compare to state-of-the-art LMs for code?}
In this RQ, we evaluate whether the use of subword units
allows the training of effective neural language models.
Specifically, we compare subword unit NLMs to standard $n$-gram
language models \cite{Hindle2012}, cache LMs \cite{Tu2014},
 state-of-the-art $n$-gram models with nested caching \cite{Hellendoorn2017},
and token-level NLMs \cite{White2015}.
Unfortunately, it was not possible to PHOG \cite{Bielik2016} because the public implementation
of PHOG does not include
the code for the first stage that generates the HOG rules.
This stage is necessary to be able to run the implementation on any dataset other than
the one used in the original paper \cite{Bielik2016}.
Their dataset is both smaller than our full dataset,
and the split is based on a within project scenario,
rather than the cross-project scenario that is the focus of our paper,
so it is not possible to address our RQs using their data.
We also do not compare to the recent pointer network augmented RNN \cite{Li2018} as it was evaluated on the dataset of \cite{Bielik2016} and  no implementation of their model is available.

As described in Section~\ref{sec:methodology}, we hypothesize
that NLMs might outperform all of the $n$-gram models
because of their ability to take larger context into account,
and
that subword unit NLMs might perform better than token level
NLMs because of the improved ability to handle OOV and rare words.
We do not report results for character-level models since as discussed in Section~\ref{sec:related} these models have not been proved to offer improvement in NLP and their use is impractical.
We also do not include results for subtoken models segmented via the heuristic in \cite{Allamanis2015a} as in preliminary experiments, we found that subtoken models were less effective than token level models, so we chose to not include them in this comparison.
Following the previous work, we evaluate the models on the Github Java
dataset (Section~\ref{sec:datasets}) using the evaluation framework
described in Section~\ref{sec:evaluation}.

\emph{RQ2. Are subword unit NLMs effectively trainable on large code corpora,
such as giga-token corpora? Does the additional training data yield significant performance improvements?} 
\label{sec:rq2Definition}
Training on a larger corpus can usually be expected to improve 
the model's performance, but this is not guaranteed, because the impact
of more data tends to have diminishing
returns, as after some point the model's performance saturates and does not
continue to improve with more data. This saturation point will be different
for different models, so it is an interesting research question to ask
whether neural LMs can make better use of large corpora than $n$-gram models,
because NLMs are more complex models that can take into account larger amounts of context.

However, training on larger data uses more resources because the amount of parameters that have to be estimated also grows with the amount of training data.
For example, $n$-gram language models need to estimate counts for every new $n$-gram that appears in the training corpus, while NLMs need to learn input and output embedding matrices whose dimension scales with the vocabulary size.
Token-level models are especially difficult to train on large corpora since they have particular difficulty with the huge vocabularies.
A huge vocabulary results in a massive model that is unable to fit in any modern GPU and would be too slow to train even with the help of softmax approximation techniques such as noise contrastive estimation \cite{Gutmann2012} and adaptive softmax \cite{Grave2016}.
Furthermore, the use of approximation techniques could potentially cause a small decrease in the model's performance.
Character level models will be too slow to train on very large training sets and impractical to use.
This problem is also especially relevant to code LMs,
because the vocabulary size for a code corpus grows much more quickly than
a natural language corpus.
For example,
the one billion word benchmark corpus for English has a vocabulary of 0.8 million words \cite{Chelba2013}, while our similar size training set for Java has a vocabulary of 6.5 million when in both cases all words are discarded with count below 3.
This huge difference in vocabulary size is also true for other programming languages.
Specifically, the training set used in our experiments for Java, C, and Python have vocabulary sizes 10.5, 8, and 13 million respectively when no vocabulary threshold is used.
Additionally, while OOV rate is only $0.32\%$ for the English corpus test set, for the Java one, it is larger than $13\%$.
The model required for a token-level code corpus of this size cannot fit in modern GPUs and is only trainable on multiple CPU threads using parallelisation and softmax approximation techniques \cite{Chelba2013}. %

\emph{RQ3. How does the performance of subword unit NLMs vary across programming languages?}
In principle the learning methods for NLMs are language agnostic;
however, the majority of studies evaluate only on Java code,
so it is an important research question to verify that current
LM techniques for code are equally effective on other programming
languages. One might hypothesize that the terseness of C
or the lack of static type information in Python would make it
more difficult to achieve good performance from an LM.
We test this hypothesis by measuring cross entropy and MRR
in the static and dynamic setting across the three language
corpora described in Section~\ref{sec:datasets}.

\emph{RQ4. Is the dynamic updating procedure effective at dynamically updating
subword unit NLMs to new projects?}
Past research has focused on the strong locality that characterises code
\cite{Hindle2012, Tu2014, Ray2016}. As a consequence, we expect new projects
to introduce many new identifiers that do not appear even in a large
cross-project corpus. For this reason, it has been shown that $n$-gram models
can benefit significantly from dynamically adapting to the test corpus, as
described in Section~\ref{sec:evaluation}. In this RQ we ask whether NLMs
can also benefit from dynamic adaptation, and whether the procedure
that we introduce in Section~\ref{sec:adaptation} is effective at dynamically
adapting NLMs to new projects. We test this hypothesis 
by comparing our dynamic adaption method for subword unit NLMs
against two advanced $n$-gram models that have been shown to benefit
from adaptation:
cache LMs \cite{Tu2014} and nested cache LMs \cite{Hellendoorn2017}.
We use the dynamic and software maintenance settings described in Section~\ref{sec:evaluation},
following \cite{Hellendoorn2017}.
A naive approach to the software maintenance setting would require
retraining the model from scratch for every file in the test corpus,
which was rightly deemed to be infeasible for NLMs by \cite{Hellendoorn2017}.
Instead, we apply our dynamic adaptation procedure from Section~\ref{sec:adaptation}, which is much more efficient because it trains for
only one epoch on each test file.

\section{Results}

\subsection{RQ1. Performance of models}

In Tables~\ref{tab:Results} and~\ref{tab:MRRResults}, we show the performance
of the different models on the static, dynamic, and maintenance settings for Java.
Note that for cross entropy, lower numbers are better, whereas for MRR, higher numbers are better. 
The entropy results for the $n$-gram models are copied from \cite{Hellendoorn2017}; these are comparable to ours because we use the same training and test split as their work. 
For MRR evaluation we used v0.1 of their model hosted in GitHub, which is the version used in their reported experiments.\footnote{The nested cache n-gram implementation can be found in \url{https://github.com/SLP-team/SLP-Core}} 
In \cite{Hellendoorn2017} the reported MRR had been calculated on the entirety of the test set).
While for their NLM baselines it was measured only on the first 1 million tokens of the test set.
For this reason we recalculated MRR for the above on the first 1 million tokens of our test set. 
The closed vocabulary NLM is our own implementation.

From the tables it can be seen that on both metrics,
our open vocabulary NLM 
has better predictive performance than any of the $n$-gram models,
even the nested cache models of \cite{Hellendoorn2017} that are designed
specifically for code.
To specifically evaluate the effect of relaxing the closed vocabulary
assumption, we compare our open vocabulary NLM to a closed vocabulary one.
The closed vocabulary NLM uses exactly the same architecture as our open vocabulary models
(single layer GRU with the same model size and hyperparameters),
but is trained on complete tokens rather than subword units.
We note that the closed vocabulary NLM that we report has much better
results than the NLM language model that is used as a baseline
in \cite{Hellendoorn2017}; this is primarily because our model incorporates
a fully connected hidden layer, and also dropout, which has been
shown to improve the performance of RNN LMs \cite{mellis17lstm}.
So our token-level NLM baseline is much harder to beat than those
reported in previous work. Even so, we find that our open vocabulary
NLM has much better predictive performance than the closed vocabulary model.
The difference in performance between the open and closed vocabulary NLMs 
is larger for the dynamic and maintenance settings than for the static setting.
We hypothesize that this is because in the open vocabulary model, dynamic adaptation
can help the model to learn patterns about OOV words in the test set; this is not
possible for a model with a closed vocabulary.

We report the performance of the open vocabulary NLMs with different vocabulary sizes, obtained after 2000, 5000, and 10000 BPE merge operations.
We see that performance is similar across the different vocabulary sizes,
indicating that a large vocabulary size is not required for good performance.

Finally, following \cite{Hellendoorn2017}, note that we cannot report results for the
nested or cache $n$-gram models on the static setting because these models make use of information from the test.
Consequently, even if we do not adapt their global model on the test set, their additional components are always adapted on it.
Also, following \cite{Hellendoorn2017} we do not report results from the closed vocabulary NLM on the maintenance setting due to the massive time and space requirements of this experiment.

Based on these results, we conclude that even when trained
on a relatively small corpus, open vocabulary NLMs are effective
models of code.
Indeed, to our knowledge, our model has state of the art
performance on this data set.

\begin{table}[tb]
\centering
\caption{Performance on next token prediction, measured by cross-entropy.}
\label{tab:Results}
\renewcommand{\arraystretch}{1.2}
\resizebox{\columnwidth}{!}{
\begin{tabular}{l@{} l c c c}
\toprule
\multicolumn{2}{l}{\textbf{Model}} & {\textbf{Static}} & {\textbf{Dynamic}} & {\textbf{Maintenance}} \\
\midrule
\multicolumn{2}{l}{Small Train}\\
\hspace{1.5em} & $n$-gram & 6.25 & 5.54 & 5.30 \\
& Nested $n$-gram \cite{Hellendoorn2017} & - & 3.65 & 2.94 \\
&  Cache $n$-gram \cite{Tu2014} & - & 3.43 & 3.32 \\
&  Nested Cache $n$-gram \cite{Hellendoorn2017}  & - & 2.57 & 2.23 \\
&  NLM (closed vocab) & 4.30 & 3.07 & -* \\
&  Open Vocab NLM (2000) & 4.90 & 2.33 & \textbf{1.46} \\
&  Open Vocab NLM (5000) & 4.78 & \textbf{2.27} & 1.51 \\
&  Open Vocab NLM (10000) & \textbf{4.77} & 2.54 & 1.60 \\
\multicolumn{2}{l}{Full Train}\\
&  Nested Cache $n$-gram \cite{Hellendoorn2017} & - & 2.49 & 2.17 \\
&  Open Vocab NLM (2000) & 3.59 & 1.84 &  \textbf{1.03} \\
&  Open Vocab NLM (5000) & 3.35 & 1.72 & 1.06 \\
&  Open Vocab NLM (10000) & \textbf{3.15} & \textbf{1.70} & 1.04 \\
\bottomrule
\end{tabular}}
\end{table}

\begin{table}[tb]
\caption{Performance of language models on suggesting the next token, 
measured by mean reciprocal rank (MRR).}
\label{tab:MRRResults}
\centering
\renewcommand{\arraystretch}{1.2}
\resizebox{\columnwidth}{!}{
\begin{tabular}{@{} l @{} l c c c}
\toprule
\multicolumn{2}{l}{\textbf{Model}} & {\textbf{Static}} & {\textbf{Dynamic}} & {\textbf{Maintenance}} \\
\midrule
\multicolumn{2}{l}{Small Train}\\
\hspace{1.5em}  & $n$-gram & 53.16\% & 56.21\% & 58.32\% \\
& Nested $n$-gram \cite{Hellendoorn2017} & - & 66.66\% & 71.43\% \\
& Cache $n$-gram \cite{Tu2014} & - & 69.09\% & 70.23\% \\
& Nested Cache $n$-gram \cite{Hellendoorn2017} & - & 74.55\% & 77.04\% \\
& NLM (closed vocab) & 62.35\% & 71.01\% & - \\
& Open Vocab NLM (2000) & 62.87\% &  76.94\% & 77.48\% \\
& Open Vocab NLM (5000) & \textbf{63.80\%} & \textbf{77.51\%} & 78.49\% \\
& Open Vocab NLM (10000) & 63.75\% & 77.32\% & \textbf{78.69\%} \\
\multicolumn{2}{l}{Full Train}\\
& Nested Cache $n$-gram \cite{Hellendoorn2017} & - & 75.02\% & 77.38\% \\
& Open Vocab NLM (2000) & 68.96\% & 78.99\% & 78.85\% \\
& Open Vocab NLM (5000) & 69.87\% & 79.88\% & 80.31\% \\
& Open Vocab NLM (10000) & \textbf{70.84\%} & \textbf{80.36\%} & \textbf{81.16\%} \\
\bottomrule
\end{tabular}}
\end{table}

\subsection{RQ2. Large Corpora}

When trained on larger corpora the performance of traditional $n$-gram models and their variations like the nested cache model gets saturated and they are unable to effectively leverage the extra information  \cite{Hellendoorn2017}. 
In contrast, our model is able to better leverage the increase in training data as shown in Tables~\ref{tab:Results} and \ref{tab:MRRResults}.
As expected the entropy of our NLM decreased significantly, by about 1.5 bits and MRR increased by about 6\% for all encoding sizes in the static scenario when trained on the full corpus.
Essentially, this means that the additional training data helps our NLM learn to synthesize identifiers from subword units better and with higher confidence.

The improvements are smaller but still exist when the model is dynamically adapted on a test project.
For all encoding sizes the models improve by 0.5 bits in entropy and by about 2 to 3\% in MRR.
In contrast, the nested cache $n$-gram model entropy decreases by less than 0.1 bits and MRR less than 0.4\%. 
Similar improvements for the nested cache $n$-gram model were also reported in \cite{Hellendoorn2017}, supporting our findings. 
From that we conclude that subword unit NLMs can utilize a large code corpus better than $n$-gram models.
However, if one lacks a model trained on large corpus or there is not enough time to train one, then satisfactory performance can still be achieved by training on a small corpus.

In addition, we note that training on giga-token or larger corpora is scalable.
Table~\ref{VRAM_Reqs} shows the VRAM requirements for training when a BPE segmentation with 2000, 5000, 10000 operations and a batch size of 32 is used.
Obviously, training time will be a lot larger than the smaller set,
but we see from the earlier results that the model's predictive performance of the open vocabulary models was stable across different vocabulary sizes.
This means that we can use a relatively moderate vocabulary size with
the open vocabulary NLM and still obtain good performance.
As training is an one-off process that does not need to be repeated and a pre-trained model can be downloaded and loaded in a GPU in a matter of seconds, this results in a real time applicable model even when trained on huge corpora.
More importantly, as reported by \cite{Hellendoorn2017} a token level NLM with a vocabulary of only 76K tokens required a few days to be trained on the smaller training corpus.
Training our model on the same data was a matter of less than 12 hours on a single GPU.

Lastly, our model can be integrated in IDEs to facilitate code completion as queries for the next token can be answered in real time and the required memory is less than that of training since batching is no longer necessary. The decreased memory requirements are illustrated in Table~\ref{completion_VRAM_Reqs} and are less than 400MBs for any of the encodings we used.

\begin{table}[tb]
\caption{Training VRAM Requirements for our BPE NLM with an encoding of 2000, 5000, and 10000 operations.}
\centering
\begin{tabular}{c c}
\hline
\textbf{BPE Ops} & {\textbf{VRAM Required}} \\
\hline
\hline
2000 & 1251 MiB\\
5000 & 1387 MiB\\
10000 & 2411 MiB\\
\hline
\end{tabular}
\label{VRAM_Reqs}
\end{table}

\begin{table}[tb]
\caption{Code completion VRAM Requirements for our BPE NLM with an encoding of 2000, 5000, and 10000 operations.}
\label{completion_VRAM_Reqs}
\centering
\begin{tabular}{c c}
\hline
\textbf{BPE Ops} & {\textbf{VRAM Required}} \\
\hline
\hline
2000 & 259 MiB\\
5000 & 297 MiB\\
10000 & 355 MiB\\
\hline
\end{tabular}
\end{table}

\subsection{RQ3. Multiple Languages}

In Tables~\ref{tab:LangEntropyStaticResults} and~\ref{tab:LangMRRStaticResults} we report the performance
of our open vocabulary NLMs on Java, C, and Python. 
Tables~\ref{tab:LangEntropyDynamicResults} and~\ref{tab:LangMRRDynamicResults} 
present the results for the dynamic setting. We see 
that performance on C and Python is at least as good as Java,
providing evidence that our methodology for training subword unit NLMs 
is indeed language agnostic.
We caution the reader to \emph{not} interpret these results as a comparison
of the programming languages as to which is more predictable,
which is more terse, etc. 
The first reason for this caution is that the training corpora have slightly different
sizes across the different languages. 
Unfortunately, it does not seem possible
to define a
fair notion of "same training set size" across programming languages, 
because tokens in one language might be more informative than others, e.g.
Python code has a larger proportion of identifiers.
Even if it were possible to do this,
different languages have different standard libraries
and are typically used to solve problems in different domains.
All of these concerns pose serious threats to validity to any attempt
to compare programming languages via language modeling, so we do not attempt
to draw such conclusions in this work.

\begin{table}[tb]
\caption{Cross-entropy in the static scenario for Java, Python, and C. Note that training set sizes vary between the languages as this is not meant to be a comparison between them.}
\label{tab:LangEntropyStaticResults}
\centering
\begin{tabular}{l@{} l c c c}
\hline
\multicolumn{2}{l}{\textbf{Model}} & {\textbf{Java}} & {\textbf{C}} & {\textbf{Python}} \\
\hline
\hline
\multicolumn{2}{l}{Small Train}\\
\hspace{1em} & Open Vocab NLM (2000) & 4.90 & 4.61 & 4.28 \\
& Open Vocab NLM (5000) & 4.78 & 4.40 & 4.03 \\
& Open Vocab NLM (10000) & 4.77 & 4.32 & 3.95 \\
\multicolumn{2}{l}{Full Train}\\
& Open Vocab NLM (2000) & 3.59 & 3.48 & 3.56 \\
& Open Vocab NLM (5000) & 3.35 & 3.43 & 3.27 \\
& Open Vocab NLM (10000) & 3.15 & 3.11 & 3.04 \\
\hline
\end{tabular}
\end{table}

\begin{table}[tb]
\caption{MRR in the static scenario for Java, Python, and C. Note that training set sizes vary between the languages as this is not meant to be a comparison between them.}
\label{tab:LangMRRStaticResults}
\centering
\begin{tabular}{l@{} l c c c}
\hline
\multicolumn{2}{l}{\textbf{Model}} & {\textbf{Java}} & {\textbf{C}} & {\textbf{Python}} \\
\hline
\hline
\multicolumn{2}{l}{Small Train}\\
\hspace{1em} & Open Vocab NLM (2000) & 62.87\% & 63.27\% & 82.07\% \\
& Open Vocab NLM (5000) & 63.80\% & 64.51\% & 81.56\% \\
& Open Vocab NLM (10000) & 63.75\% & 66.09\% & 81.66\% \\
\multicolumn{2}{l}{Full Train}\\
& Open Vocab NLM (2000) & 68.96\% & 67.19\% & 82.92\% \\
& Open Vocab NLM (5000) & 69.87\% & 67.64\% & 83.18\% \\
& Open Vocab NLM (10000) & 70.84\% & 70.35\% & 84.31\% \\
\hline
\end{tabular}
\end{table}

\subsection{RQ4. Dynamic Adaptation}

Finally, to evaluate the effect on the dynamic adaptation method for our subword unit NLMs, consider again the results in Tables~\ref{tab:Results} and~\ref{tab:MRRResults}. As \cite{Hellendoorn2017} point out, it is 
 straightforward to adapt an $n$-gram LM, because we can simply add
 and remove counts.  Indeed, we see that all of the advanced $n$-gram models in the dynamic and maintenance settings perform better than any of the NLM
 models in the static setting.
This result holds both for the small train set and for the full train set.
In other words, the improvement due to dynamic adaptation is greater
than the improvement due to an NLM. 
Once we apply the dynamic adaptation method to our Open Vocabulary NLM, however,
then the picture changes.
With dynamic adaptation, 
our model achieves better cross-entropy than the current state-of-the-art \cite{Hellendoorn2017}.
From this we conclude that our dynamic adaptation method is indeed
effective at fine-tuning a global subword unit NLM to a specific test project.

We note that evaluating NLMs on this scenario was previously deemed infeasible since multiple models had to be created each trained on the entirety of the test set minus one file. Nevertheless, the small size of our model allowed the experiments for this scenario to be completed in  a few days. We achieved this by training each model only on information from the same project.
 For large test projects, we first split them into multiple partitions and for each one we trained a model on the rest. All files from the same partition can then load this model and need to only train on other files from the same partition. This strategy offered considerable speed gains. 
Furthermore, the experiment could be sped up significantly as parallelization is fairly easy and both memory and computation requirements are fairly small, thus achievable even with a single GPU.

\begin{table}[tb]
\caption{Cross-entropy in the dynamic scenario for Java, Python, and C. Note that training set sizes vary between the languages as this is not meant to be a comparison between them.}
\label{tab:LangEntropyDynamicResults}
\centering
\begin{tabular}{l@{} l c c c}
\hline
\multicolumn{2}{l}{\textbf{Model}} & {\textbf{Java}} & {\textbf{C}} & {\textbf{Python}} \\
\hline
\hline
\multicolumn{2}{l}{Small Train}\\
\hspace{1em} & Open Vocab NLM (2000) & 2.33 & 1,79 & 3.28 \\
& Open Vocab NLM (5000) & 2.27 & 1.74 & 3.16 \\
& Open Vocab NLM (10000) & 2.54 & 1.72 & 3.13 \\
\multicolumn{2}{l}{Full Train}\\
& Open Vocab NLM (2000) & 1.84 & 1.69 & 2.83 \\
& Open Vocab NLM (5000) & 1.72 & 1.59 & 2.67 \\
& Open Vocab NLM (10000) & 1.70 & 1.56 & 2.52 \\
\hline
\end{tabular}
\end{table}

\begin{table}[tp]
\centering
\caption{MRR in the dynamic scenario for Java, Python, and C. Note that training set sizes vary between the languages as this is not meant to be a comparison between them.}
\label{tab:LangMRRDynamicResults}
\begin{tabular}{l@{} l c c c}
\hline
\multicolumn{2}{l}{\textbf{Model}} & {\textbf{Java}} & {\textbf{C}} & {\textbf{Python}} \\
\hline
\hline
\multicolumn{2}{l}{Small Train}\\
\hspace{1em}& Open Vocab NLM (2000) & 76.94\% & 72.22\% & 86.27\% \\
& Open Vocab NLM (5000) & 77.51\% & 72.76\% & 86.31\% \\
& Open Vocab NLM (10000) & 77.32\% & 73.31\% & 86.28\% \\
\multicolumn{2}{l}{Full Train}\\
& Open Vocab NLM (2000) & 78.99\% & 72.49\% & 86.51\% \\
& Open Vocab NLM (5000) & 79.88\% & 73.63\% & 86.85\% \\
& Open Vocab NLM (10000) & 80.36\% & 74.32\% & 87.06\% \\
\hline
\end{tabular}
\end{table}

\section{Conclusions}

We have presented a new open-vocabulary neural language model for source code.
By defining the model on subword units, which are character subsequences of
tokens, the model is able to handle neologisms, that is, new identifier names which have not 
appeared in its training data, while keeping the size of the model
relatively small.
We are able to train a neural
language model on over one billion tokens of code, a data set over a hundred
times larger than had been used for previous neural LMs for code.
On the problem of predicting the next token, the resulting model
outperforms recent state-of-the-art models based on adding nested caches
to $n$-gram language models.
We hope that the simplicity of our model will allow advances in deep learning for code by allowing the implementation of more complex architectural ideas such as
attention \cite{Bahdanau2014,transformer}. Also, improved language models for code have the potential
to enable new tools for aiding code readability \cite{Allamanis2014}, program repair 
\cite{Ray2016,Campbell2014,Gupta17,Bhatia2016}, program synthesis \cite{gulwani2017program} and translation between programming languages \cite{Karaivanov2014,Nguyen2013b}.
Finally, the general technique of using subword units is not limited to language modeling, but
can easily be incorporated into many neural models of code tokens.
Therefore, we hope that this idea could have broad application throughout software engineering,
such as in models to suggest readable function and class names \cite{Allamanis2015a}, summarizing source code \cite{Iyer2016, Allamanis2016}, predicting bugs \cite{Pradel2018}, detecting code clones \cite{White2016}, comment generation \cite{Hu2018}, fixing syntactic errors \cite{Kruthiventi2015}, and variable de-obfuscation \cite{Bavishi2018}.

\bibliographystyle{ACM-Reference-Format}
\bibliography{BPEPaperBib}

\end{document}